# Transient THz conductivity of silicon with optical pumping above and below the second indirect transition

F. Novelli,[1] J. A. Davis,[1,a]

[1] *Centre for Quantum and Optical Science, Swinburne University of Technology, Hawthorn, Victoria, Australia, 3122*

[a] jdavis@swin.edu.au

Here we perform a series of time-resolved experiments where a 100 fs pump pulse is tuned between 528 nm and 555 nm, across the second indirect gap of intrinsic silicon at ~540 nm which involves electrons in a higher-lying conduction band with minimum at the L point. The photo-injected carriers, after inter- and intra-band relaxations are complete, are subsequently probed with high-field single-cycle terahertz radiation. When the energy of the pump pulses exceeds the second indirect gap, the probed terahertz absorption decreases by a factor 2.7±0.2. We suggest that this dramatic change could be due to the different phonon populations obtained when the carriers undergo the L→X inter-band scattering, instead of just cooling within the X-valley.

**I. INTRODUCTION**

Owing to their multiple technological applications, the transport properties of electrons and holes in semiconductors have been extensively studied in the last decades. The development of intense pulsed high-field sources, such as the ones based on tilted-front terahertz (THz) generation [1], allowed to study the non-thermal carrier transport phenomena in crystalline materials with a high temporal resolution [2-9]. A variety of intriguing dynamical effects have been found in different semiconductors. For example experiments employing intense THz pulses demonstrated impact ionisation in InSb [2] and InAs [3]; ultrafast saturable absorption of GaAs, GaP, and Ge [4]; band nonparabolicity in doped InGaAs [5]; self-phase modulation of single-cycle THz pulses in n-GaAs [6]; band-gap softening via the dynamical Franz-Keldysh effect [7]; emission of high-harmonics [8]; and coherent, quantum-kinetic ballistic motion of the electrons in the conduction band [9].

In addition, THz-pump THz-probe experiments revealed a drop in the absorption of the probe, or bleaching, at pump-probe overlap in doped InSb [2], InAs [3], GaAs [10-12], Ge [13], and Si [13]. In a simplified view these results can be understood with the intense THz pump pulses exciting carriers into higher energy valleys with larger effective masses and thus reduced mobility and absorption. After the pump-probe overlap, the photo-excited carriers relax down to the bottom (top) of the lowest valley in the conduction (valence) band within few hundreds femtoseconds [14,15] and the absorption recovers its initial value. While this explanation accounts for most of the observations, it is inadequate to explain the response of silicon. In silicon, in fact, the second indirect gap involves electronic excitations in the L valley which is characterised by a smaller effective mass [14] than the lowest-energy X valley and, hence, one would simply expect an absorption increase at the overlap of THz-pump and THz-probe pulses followed by a full relaxation within ~1 ps. On the contrary, the frequency-integrated THz pump-probe response of Si displays a drop at pulse overlap followed by a bi-exponential relaxation with a fast component ~1 ps long, and a slow component with decay constant of ~24 ps (see Fig. 3 in Ref. [13] and analysis therein). While the drop at time zero could still be qualitatively explained by a substantial high-field mobility saturation [16,17] and/or band nonparabolicity [5], the origin of the longer relaxation remains unclear.

Here we perform a series of pump-probe experiments combining visible pump pulses with frequency-integrated THz probes. The visible pulses are tuned across the energy of the second indirect gap of undoped silicon and the THz peak field is acquired as a function of the pump-probe delay. When the photon energy is larger than the second indirect gap of silicon the transient THz absorption decreases. We suggest that carriers initially excited in the L conduction valley relax to the lowest states around the X point by creating a phonon distribution which scatters faster with the carriers, resulting in a reduced mobility and in a smaller photo-induced terahertz absorption. The recovery of the THz absorption in the measurements here and in the previous THz-pump THz-probe experiments is then limited by relaxation of the phonon population.

## II. RESULTS

In order to generate pulses at THz and visible frequencies we use a Ti:Sa amplified laser source emitting 0.4 mJ/pulse with 1 kHz repetition rate and wavelength 800 nm. The source is split, with half used for THz generation and detection and half going to a two stage OPA where tunable pulses in the visible range (555-528 nm central wavelength, about 7 nm FWHM bandwidth) are generated by sum generation of the idler with residual Ti:Sa pump. The THz fields are generated by tilted-front optical rectification of the amplified pulses in 0.6% MgO doped stoichiometric $LiNbO_3$ [1], with an energy conversion efficiency of about $2·10^{-4}$. The fields are detected by electro-optical sampling in a 0.5 mm thick <110> ZnTe crystal [18].

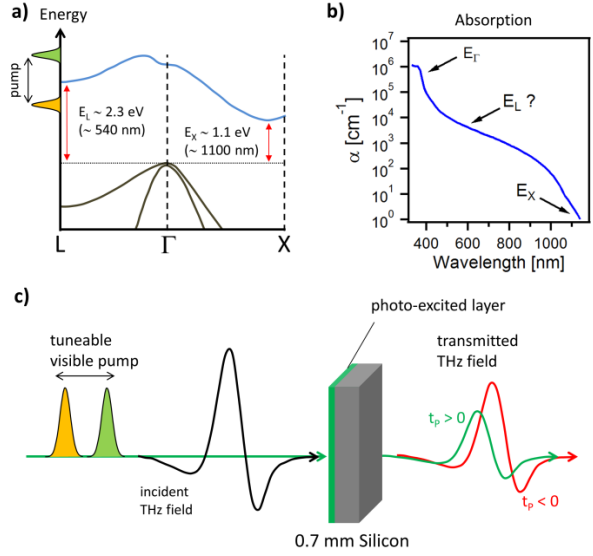

*Figure 1. a) Cartoon of the electronic bands in bulk silicon along the ΓL and ΓX directions. b) Absorption spectrum of intrinsic silicon, with the position of the first (second) indirect gap $E_X$ ($E_L$) indicated (data from Table 1 in Ref.[19]). c) Cartoon of the two-pulse experiment.*

The band structure of silicon is sketched in Fig.1a along the ΓL and ΓX directions. The well-known first indirect gap with $\lambda_X$~1100 nm ($E_X$~1.1 eV) from the Γ point in the valence band to the conduction band valley close to the X point is evident. The second lowest indirect gap at $\lambda_L$~540 nm ($E_L$~2.3 eV) [20-22] involves the conduction band minimum located at the L point. In the equilibrium optical response of silicon (Fig.1b) [14,19] there is, however, no apparent signature of the excitation of carriers across this gap. Here we tune the excitation pump wavelength across this second indirect gap and probe the photo-excited carriers with terahertz pulses.

The visible pump photo-excites a thin layer at the silicon surface (green in Fig.1c) which modifies the transmitted, and subsequently detected, THz field (as sketched with the $t_P > 0$ green field on the right side of Fig.1c). The THz spot (intensity, ~2 mm FWHM) is about half as large as the pump spot (~4 mm FWHM at all wavelengths used) thus minor artefacts in the THz spectrum can be present [23,24]. In the following, however, we do not discuss specific spectral changes, and instead focus on variations of the THz peak field (which is effectively the integrated spectrum). The THz field is measured by delaying simultaneously both sampling and pump pulses so that the delay between them remains fixed. In this way each point of the THz field is detected at the same time delay with respect to the pump arrival. In the following we will refer to the fixed delay between sampling and pump beams as "pump delay" ($t_P$), while "sampling delay" (t) will be used to refer to the temporal shift imposed to both pump and sampling pulses simultaneously. The directions of the visible and of the THz pulses make a small angle at the sample position, resulting in a time resolution of about 1 ps at pump-probe overlap [25].

In Fig.2 we show the THz peak variation, $\Delta E_{max} = E(t=0, t_P) - E_{eq}(t=0)$, as a function of the laser power for the longest central pump wavelength used, $\lambda_P = 555$ nm $> \lambda_L$. The pump fluence, $\Phi$, is varied from 1.1 µJ/cm$^2$ to 6.5 µJ/cm$^2$, and the pump delay up to 80 ps. The transient THz response of silicon is linear with $\Phi$ over this range and reaches a plateau within less than 2 ps after the pump arrival and remains roughly constant up to 80 ps.

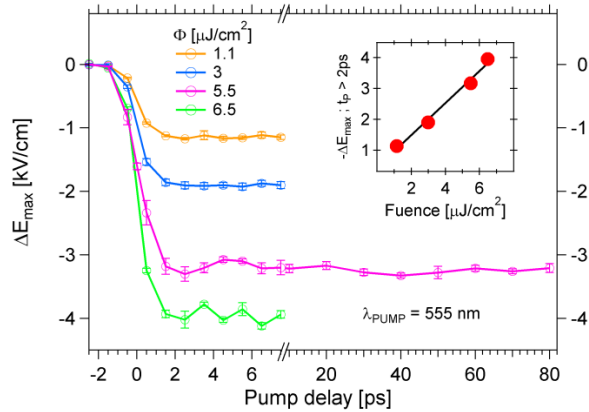

*Figure 2. a) Pump-induced variation of the THz pulses at the peak field, $\Delta E_{max}$, for different pump intensities. The central wavelength of the pump was 555 nm with a FWHM bandwidth of 7 nm. The inset plots the average of $-\Delta E_{max}$ for positive pump delays (from +2 ps to +7 ps) as a function of pump fluence. Error bars are ±1 standard deviations (sd) calculated from 8 measurements.*

We now vary the excitation central wavelength from $\lambda_P=555$ nm to $\lambda_P=528$ nm. Additional measurements at $\lambda_P=560$ nm and $\lambda_P=565$ nm are not shown, but are fully consistent with the results obtained for $\lambda_P=555$ nm and $\lambda_P=550$ nm. We set the pump fluence on the low side of the linear response regime reported in Fig.2 ($\Phi=2.4$ µJ/cm$^2$); vary the pump delay $t_P$ up to 7.5 ps and the sampling times t to values around the THz peak.

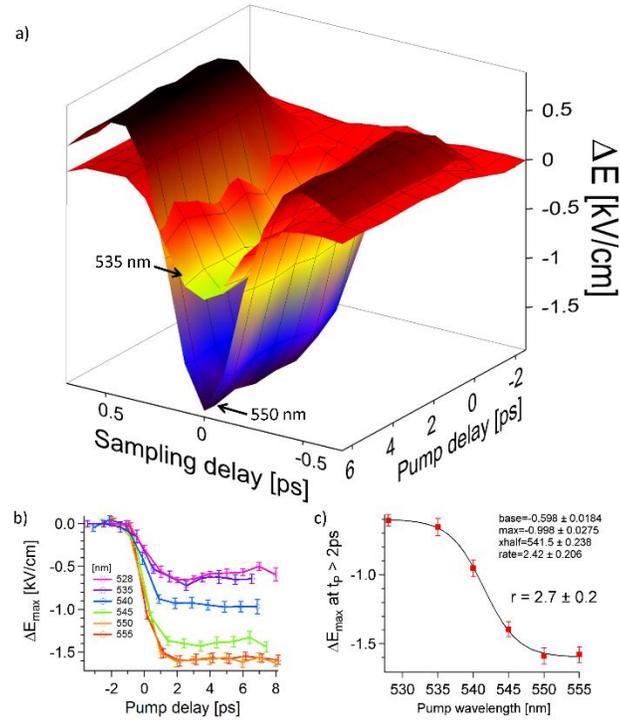

*Figure 3. a) Variation of the transmitted terahertz field as a function of sampling and pump delay times for $\lambda_P$=550 nm and $\lambda_P$=535 nm. b) Pump-induced variation of the THz peak field as a function of pump delay for pump wavelengths from 555 nm to 528 nm. c) Average pump-induced THz response at pump delays larger than 2 ps (red squares) as a function of pump wavelength. The solid line is the fit to a step function: base+max/{1+exp[(xhalf-x)/rate]}. r is defined as (base+max)/max and quantifies the size of the wavelength-dependent effect. Error bars in b) and c) are ±1 sd calculated from 15 measurements.*

Results at $\lambda_P$=535 nm and $\lambda_P$=550 nm are shown in Fig.3a. Here the pump-induced variation of the transmitted THz field is shown as a function of the pump and sampling delays. Even though the equilibrium optical properties of undoped silicon are similar at 535 nm and 550 nm [14,19], the transient THz response of the system is strongly diminished for $\lambda_P$=535 nm. The transmitted peak field is shown in Fig.3b as a function of pump delay for the different wavelengths used; namely 528 nm, 535 nm, 540 nm, 545 nm, 550 nm, and 555 nm with bandwidth about 7 nm FWHM. The average of $\Delta E_{max}$ at positive pump delays $t_P > 2$ ps as a function of the pump wavelength is shown in Fig.3c. The photo-induced average absorption of the transmitted terahertz fields diminishes at the shorter wavelengths used. We can estimate the wavelength-dependent response by fitting the peak field variation with a step function (see Fig.3c and its caption). This fitting procedure allows to estimate the parameter r=2.7±0.2 accounting for the different responses when the pump wavelength is above or below $\lambda_L$, i.e., the transient THz response is ~2.7 times larger when the pump wavelength is smaller than $\lambda_L$.

## III. DISCUSSION

Optical-pump terahertz-probe measurements have been performed on a variety of materials such as semiconductors [26,27], molecular-based photovoltaics [28], graphene [29], and others [30,31] in order to obtain their transport properties via non-contact probes. For simple materials, whose optical properties can be described with a Drude model, the photo-induced THz absorption is proportional to the weighted sum of the carriers mobilities [30]. Each mobility μ, in turn, is given by the ratio e·τ/m$^*$ with e elementary charge, m$^*$ effective mass, and τ scattering time of the carriers. If the effective mass increases, as can happen if the carriers are injected into a higher lying valley [2,3,10-12] or if they are driven into non-parabolic portions within the same band [5], the mobility decreases. Conversely if the carrier-phonon scattering time varies, for example as the result of a changed phonon population, one expects a different mobility.

By exciting silicon with light pulses at wavelengths between the first and the second indirect gaps, corresponding to the measurements for $\lambda_P$ > 545 nm (Fig.3), one expects that electrons and holes are injected at some high energy points in the X and Γ valleys, respectively. After a fast intraband relaxation shorter than ~200 fs [14,15], electrons (holes) occupy the lowest energy states of the X (Γ) valley in the conduction (valence) band where they remain for hundreds of picoseconds [15,25]. At positive pump-probe delays (exceeding ~200 fs) the THz field probes these relaxed carriers and, because these bands were effectively empty prior to the pump arrival, detects a decreased THz transmission.

When the energy of the visible laser pulses exceeds the threshold of the second indirect gap ($\lambda_P$ < 540 nm), electrons are excited in both the X and L valleys of the conduction band. Now a fast inter-band L→X carrier relaxation also takes place which, for a minimum along the L direction in a homopolar semiconductor such as silicon, is dominated by nonpolar phonon scattering [14,22]. Detailed photoemission experiments have been performed in silicon by Ichibayashi et al. [22], showing that the L→X intervalley relaxation occurs with a characteristic time of 180 fs through electron-phonon scattering involving ~30 meV LA modes. After this L→X inter-valley relaxation is complete, the electrons incur intra-band relaxation within the X valley of the conduction band with the characteristic ~200 fs timescale. Subsequently, the THz field probes these electrons close to X point together with the holes at around Γ. In this simplified view, the main difference between the $\lambda_P$ < 540 nm and the $\lambda_P$ > 545 nm cases is that the L→X inter-valley scattering is required for the former case only.

Here we suggest that the different phonon population generated through the L→X relaxation might explain the dramatic step in transient THz absorption as the pump wavelength crosses the second indirect gap energy, as well as the long lived signal seen in previous THz-pump THz-probe experiments [13]. Based on the preceding discussion, one would expect that the electron (hole) populations probed by the THz pulse more than 2 ps after the pump pulse should be very similar for each of the pump wavelengths studied here, since the absorption spectrum is roughly flat in this region, the photon density in the pump pulse is kept constant and the carrier relaxation to the conduction (valence) band minimum (maximum) is expected to be fast and efficient. However, the phonon populations after this relaxation should be different if L→X intervalley scattering has occured (which is possible for $\lambda_P < 540$ nm) or not ($\lambda_P > 545$ nm). The different phonon populations may then affect the carrier mobility due to different carrier-phonon scattering rates, leading to a difference in the transient THz absorption. This interpretation can also qualitatively account for the persistent signal in previous THz-pump-probe experiments (the slower ~24 ps decay constant) [13], whereby the different phonon population generated when the intense THz pump drives the electrons into the L-valley would require tens of picoseconds to relax via phonon-phonon scattering processes [15].

## VI. CONCLUSIONS AND PERSPECTIVES

In conclusion we performed time-resolved experiments by combining optical laser pulses and strong single-cycle THz fields on undoped silicon. When the ultrafast optical pulses are tuned above the energy threshold of the second indirect gap, involving excitations of electrons into the L valley, the transmitted THz fields are dramatically affected. Our results may be simply explained by the fast interband L→X electron relaxation resulting in different phonons populations with decreased carrier-phonon scattering time, leading to reduced carrier mobility and hence THz absorption. We suggest that this approach is able to reveal an excitation threshold that is hidden in the equilibrium optical response, which could be of significant benefit in identifying the positions of the energy gaps in other, more complex solid state systems [32-34].

## ACKNOWLEDGMENTS

We thank the Australian Research Council for funding (DP130101690). FN acknowledges funding from the Swinburne University of Technology Early Research Career Scheme 2014.